\documentclass{ws-ijmpd}
\usepackage[super,compress]{cite}
\usepackage{amsmath}
\usepackage{amsfonts}
\usepackage{graphicx}
\usepackage{dcolumn}

\begin{document}
\newcommand{\beq}{\begin{equation}}
 \newcommand{\eeq}{\end{equation}}
 \newcommand{\bea}{\begin{eqnarray}}
 \newcommand{\eea}{\end{eqnarray}}
 \newcommand{\beaa}{\begin{eqnarray*}}
 \newcommand{\eeaa}{\end{eqnarray*}}
 \newcommand{\Lhat}{\widehat{\mathcal{L}}}

\markboth{Zhang, et al.} {Model-independent constraint on the Hubble
constant}

\title{A model-independent constraint on the Hubble constant with gravitational waves from the Einstein Telescope}

\author{Sixuan Zhang}
\address{Department of Astronomy, Beijing Normal University, Beijing 100875, China;}

\author{Shuo Cao$^*$}
\address{Department of Astronomy, Beijing Normal University, Beijing 100875, China;\\
caoshuo@bnu.edu.cn}

\author{Jia Zhang}
\address{School of Physics and Electrical Engineering, Weinan Normal
University, Shanxi 714099, China;}

\author{Tonghua Liu}
\address{Department of Astronomy, Beijing Normal University, Beijing 100875, China;}

\author{Yuting Liu}
\address{Department of Astronomy, Beijing Normal University, Beijing 100875, China;}

\author{Shuaibo Geng}
\address{Department of Astronomy, Beijing Normal University, Beijing 100875, China;}

\author{Yujie Lian}
\address{Department of Astronomy, Beijing Normal University, Beijing 100875, China;}

\maketitle

\begin{history}
\received{Day Month Year}
\revised{Day Month Year}
\end{history}

\begin{abstract}

In this paper, we investigate the expected constraints on the Hubble
constant from the gravitational-wave standard sirens, in a
cosmological-model-independent way. In the framework of the
well-known Hubble law, the GW signal from each detected binary
merger in the local universe ($z<0.10$) provides a measurement of
luminosity distance $D_L$ and thus the Hubble constant $H_0$.
Focusing on the simulated data of gravitational waves from the
third-generation gravitational wave detector (the Einstein
Telescope, ET), combined with the redshifts determined from
electromagnetic counter parts and host galaxies, one can expect the
Hubble constant to be constrained at the precision of $\sim 10^{-2}$
with 20 well-observed binary neutron star (BNS) mergers. Additional
standard-siren measurements from other types of future
gravitational-wave sources (NS-BH and BBH) will provide more
precision constraints of this important cosmological parameter.
Therefore, we obtain that future measurements of the luminosity
distances of gravitational waves sources will be much more
competitive than the current analysis, which makes it expectable
more vigorous and convincing constraints on the Hubble constant in a
cosmological-model-independent way.

\end{abstract}


\ccode{PACS numbers:}

\section{Introduction} \label{introduction}

The Hubble constant $H_0$, which illustrates the expansion rate of
the universe today, plays a significant role in the deep
understanding of fundamental physics questions \cite{Weinberg13}.
Therefore, precise and accurate measurement of the Hubble constant
is one of the most fundamental issues influencing our understanding
of the Universe. Although multiple paths to independent estimates of
$H_0$ have been accessed by many astrophysical probes - in
particular the observations of type Ia supernovae (SNe Ia) and the
first acoustic peak location in the pattern of anisotropies of the
Cosmic Microwave Background Radiation (CMBR) - two issue should be
reminded. First of all, the Hubble constant cannot be constrained
directly from CMB observations, e.g. the latest Planck 2015 results
\cite{Ade16}, but must be inferred by assuming a pre-assumed
cosmological model (the standard $\Lambda$CDM model). It was found
in \cite{Cao17} that many parameters (i.e., the matter density
parameter $\Omega_m$) become degenerate with the Hubble constant: a
high value of $\Omega_m$ will lead to a low value of $H_0$.

When relaxing the $\Lambda$CDM assumption by introducing an exotic
source of matter with negative net pressure, the so-called dark
energy to explain cosmic acceleration, the strong degeneracy between
various cosmological parameters (such as the cosmic equation of
state $w=p/\rho$, or the interaction term between dark matter and
dark energy) and the the Hubble constant was also noticed and
discussed in \cite{Cao11,Cao13,Cao15,Chen15,Pan15}. Second,
alternative methods of deriving the Hubble constant from
cosmological-model-independent probes, focus on the luminosity
distance $D_L(z)$ using SNe Ia as standard candles at lower
redshifts \cite{Freedman17} and the time-delay distance $D_{\Delta
t}$ using time delays of strong lensing systems as standard rulers
\cite{Pan16,Wong19}. These results showed that recent determinations
of $H_0$, from the Supernovae $H_0$ for the Equation of State of
Dark Energy (SH0ES) collaboration \cite{Freedman17} and a joint
analysis of six gravitationally lensed quasars with measured time
delays \cite{Wong19}, are in strong tension with with the Planck CMB
measurements. The debate about the discrepancy between the Hubble
constant measured locally and the value inferred from the Planck
survey, has kept the discussion about a local underdensity alive
\cite{Riess19}. Therefore, such tension may force the rejection of
the standard $\Lambda$CDM model or indicate new physics incorporated
into cosmology.

However, it is worth noting that all of these $H_0$ measurements
performed through electromagnetic(EM) radiations. Gravitational wave
offers an independent method of determining $H_0$ and resolving the
$H_0$ discrepancy \cite{Schutz86}. The inspiraling and merging
compact binaries consisting of neutron stars (NSs) and black holes
(BHs), can be considered analogously as the supernovae (SNe)
standard candles, namely the standard sirens. The most
well-established method for measuring $H_0$ is through the Hubble
law, based on the observations of the local ¡°Hubble flow¡± velocity
of a source and the distance to the source in the local Universe.
Gravitational wave signals from inspiraling binary systems are
``standard sirens" in that the absolute value of their luminosity
distances, and thus the distances to GWs in the Hubble flow can be
determined, and therefore can be used to infer $H_0$ independent of
any other distance ladders: standard sirens are self-calibrating.
The local ¡°Hubble flow¡± velocity is typically obtained via the
identification of electromagnetic counterpart and the host galaxy.
The breakthrough took place with the first direct detection of
GW170817 in both gravitational waves and electromagnetic waves
\cite{Abbott16}, which has opened an era of gravitational-wave
multi-messenger astronomy. \cite{Abbott17} determined the Hubble
constant to be $H_0=70.0^{+12.0}_{-8.0}$ km/s/Mpc, which is well
consistent with the currently existing measurements (CMB and SNe
Ia). In the past years, many papers have studied the possibility of
the GW as standard
sirens~\cite{Holz05,MacLeod08,Sathyaprakash10,Zhao11,Cai15,Qi19a,Qi19b},
and there are also estimations on the constraint ability of the
Hubble constant by the simulated GW data
\cite{Nissanke10,Taylor12,Cai17}. Following this direction,
extensive efforts have been made to use simulated GW data to place
constraints on this important cosmological parameter \cite{Chen18},
which showed that the constraint ability of GWs is much better than
the traditional probes (with a precision of approximately two
percent), if hundreds of GW events have been observed by LIGO and
Virgo within five years.

Inspired by the previous work \cite{Chen18}, in this paper we
explore the ability of the gravitational wave detections of the
Einstein Telescope(ET) to constrain the Hubble constant in a
model-independent way based on the Hubble law. More importantly, the
third-generation ground-based detector, i.e. Einstein telescope
(ET), will be ten times more sensitive in amplitude than the
advanced ground-based detectors, covering the frequency range of
1-$10^4$ Hz. Therefore, $10^4-10^5$ GW events will be detected by ET
per year and about one of a thousand events will locate in the
low-redshift range of $[0,0.1]$ \cite{ET}. This expected
considerable number of low-redshift GW events implies that it is
possible to use these systems for estimating the Hubble constant, by
combining the measurements of the sources' redshitfs from, for
example, the electromagnetic (EM) counterpart. In this paper, we
explore the ability of the gravitational wave detections to
constrain $H_0$. As result, we obtain that future results from GWs
will be much more competitive with current limits from current
analyses. The paper is organized as follows. In Section II we
describe the methodology used in our work. The simulated GW data and
the error estimation of the standard siren measurements are
presented in Section III. In Section IV, we present the constraints
these data put on the Hubble constant. Finally, the conclusions and
discussions are presented in Section V. Throughout this paper, the
Hubble constant $H_0=70.0$km/s/Mpc from the latest GW observations
(GW170817) is taken for Monte Carlo simulations in our analysis.

\section{Methodology}

We assume that in the homogeneous and isotropic universe, its
geometry can be described by the
Friedmann-Lema$\hat{\i}$tre-Robertson-Walker (FLRW) metric
\begin{equation}
ds^2=-dt^2+\frac{a(t)^2}{1-kr^2}dr^2+a(t)^2r^2d\Omega^2,
\end{equation}
where $t$ is the cosmic time, $a(t)$ is the scale factor whose
evolution depends on the matter and energy contents of the universe,
while $k$ represents the spatial curvature. $k=+1, -1, 0$
corresponds to closed, open, and flat universe, respectively and is
related to the curvature parameter as $\Omega_k=-k/H_0^2$. In the
framework of FLRW metric, at nearby distances the mean expansion
rate of the Universe is well approximated by the expression
\cite{Hubble29}
\begin{equation}
v_H=H_0 D_H,
\end{equation}
where $v_H$ is the local ¡°Hubble flow¡± velocity of a source, $D_H$
is the Hubble distance to the source (all cosmological distance
measures, such as luminosity distance, co-moving distance and
angular diameter distance can not be distinguished at low
redshifts). In this case, the exact value of other cosmological
parameters (such as the matter density parameter $\Omega_m$, the
cosmic equation of state $w$) is not our concern, since they are
similarly insensitive to the distance measurements. In a closed
universe, this linear redshift-distance relation is usually
expressed in the form of $cz=H_0 D_H$, where $c$ is the speed of
light and $z$ is the redshift of the galaxy \cite{Hubble29}.
However, with the exception of cosmological models in which the
Hubble parameter is a constant at higher redshifts, the Hubble law
is linear only for low redshifts ($z \ll 1$) \cite{Harrison93}. When
it comes to a higher redshift, such approximation can lead to a
significant error in measuring $D_H$ and one needs to consider the
relativistic correction of the approximation \cite{Hogg99},
$1+z=\sqrt{({1+\frac{v}{c}}/{1-\frac{v}{c}}})$. With this
correction, the corrected Hubble law can be rewritten as
\cite{Carroll07}
\begin{equation}\label{18}
\frac{(1+z)^2-1}{(1+z)^2+1}c=H_0D_H.
\end{equation}
More specifically, by taking the relativistic correction form, it is
estimated that the observable distances (luminosity distance,
angular diameter distance, etc.) will differ from the Hubble
distance by less than $5\%$ (when $z\leq 2$) \cite{Carroll07}. In
this paper, we choose to implement a stringent redshift criterion
when the relativistic correction is considered (i.e, $z<0.1$).

In order to calculate $H_0$, on the one hand, we must also measure a
redshift for each binary merger. Throughout, we take the redshift
$z$ to be the peculiar-velocity-corrected redshift, i.e., the
redshift of the source if it is located in the Hubble flow.  It
should be noted two different cases will be considered in this work:
the GW sources are caused by binary merger of a neutron star with
either a neutron star or black hole, which can generate an intense
burst of $\gamma$-rays (SGRB) with measurable source redshift, or
the redshift information either comes from a statistical analysis
over a catalogue of potential host galaxies, when the GW sources are
caused by compact binaries consisting of black holes (BHs).

On the other hand, different from the luminosity distance
measurements from EM observations, the GW signal from a compact
binary system can provide the measurement in another way: through
its dependency on the amplitude of the GW event and the so-called
chirp mass of the binary system, which can be measured from the GW
signal¡¯s phasing \cite{Schutz86}. More specifically, the GW signal
from each detected binary merger (with component masses $m_1$ and
$m_2$) provides a measurement of $D_L$, which can be directly
inferred from the amplitude
\begin{eqnarray}\label{eq8}
\nonumber \mathcal{A}&=&\frac{1}{D_L}\sqrt{F_+^2(1+\cos^2(\iota))^2+4F_{\times}^2\cos^2(\iota)}\\
&\times&\sqrt{5\pi/96}\pi^{-7/6}\mathcal{M}_c^{5/6}
\end{eqnarray}
of Fourier transform of the strain $h(t)$ of GW signal
\begin{equation}\label{eq7}
\mathcal{H}(f)=\mathcal{A}f^{-7/6}\exp[i(2{\pi}ft_{0}-{\pi}/4+2{\psi}(f/2)-{\phi_{2.0}})],
\end{equation}
where $t_0$ is the epoch of the merger, while the definitions of the
functions $\psi$ and $\varphi_{(2.0)}$ can be found in
\cite{Zhao11}. In the transverse-traceless (TT) gauge, the strain
can be written as the linear combination of the two polarization
states
\begin{equation}\label{eq4}
h(t)=F_{\times}(\theta,\phi,\psi)h_{\times}(t)+F_{+}(\theta,\phi,\psi)h_{+}(t)
\end{equation}
where $h_{\times}$ and $h_{+}$ are the two independent components of
the GW tensor, $F_{\times}$ and $F_{+}$ are the beam pattern
functions, $\psi$ denotes the polarization angle, and ($\theta,
\phi$) are the location angles of the source in the sky, which
describes the location of the source relative to the detector. The
exact forms of pattern functions for ET are given by \cite{Zhao11}
\begin{align}
F_+^{(1)}(\theta, \phi, \psi)=&~~\frac{{\sqrt 3 }}{2}[\frac{1}{2}(1 + {\cos ^2}(\theta ))\cos (2\phi )\cos (2\psi ) \nonumber\\
                              &~~- \cos (\theta )\sin (2\phi )\sin (2\psi )],\nonumber\\
F_\times^{(1)}(\theta, \phi, \psi)=&~~\frac{{\sqrt 3 }}{2}[\frac{1}{2}(1 + {\cos ^2}(\theta ))\cos (2\phi )\sin (2\psi ) \nonumber\\
                              &~~+ \cos (\theta )\sin (2\phi )\cos (2\psi )].
\label{equa:F}
\end{align}
and the other two interferometer's antenna pattern functions are
$F_{+,\times}^{(2)}(\theta, \phi, \psi)=F_{+,\times}^{(1)}(\theta,
\phi+2\pi/3, \psi)$ and $F_{+,\times}^{(3)}(\theta, \phi,
\psi)=F_{+,\times}^{(1)}(\theta, \phi+4\pi/3, \psi)$, since the
three interferometers of the ET are arranged in an equilateral
triangle. We can define the chirp mass $\mathcal{M}_c=M \eta^{3/5}$
and its corresponding observational counterpart as
$\mathcal{M}_{c,\rm obs}=(1+z)\mathcal{M}_{c,\rm phys}$ (with the
total mass $M=m_1+m_2$ and the symmetric mass ratio
$\eta=m_1m_2/M^2$). $\iota$ is the angle between inclination of the
binary system's orbital angular momentum and line of sight. One
should note that, from observational point of view, the maximal
inclination is about $\iota=20^\circ$ and averaging the Fisher
matrix over the inclination $\iota$ with the limit $\iota<20^\circ$
is approximately equivalent to taking $\iota=0$. Therefore, one can
take  $\iota=0$ for simplicity, as argued in \cite{Cai18} and
references therein.

\begin{figure*}\label{fig1}
\centering
\includegraphics[width=0.6\hsize]{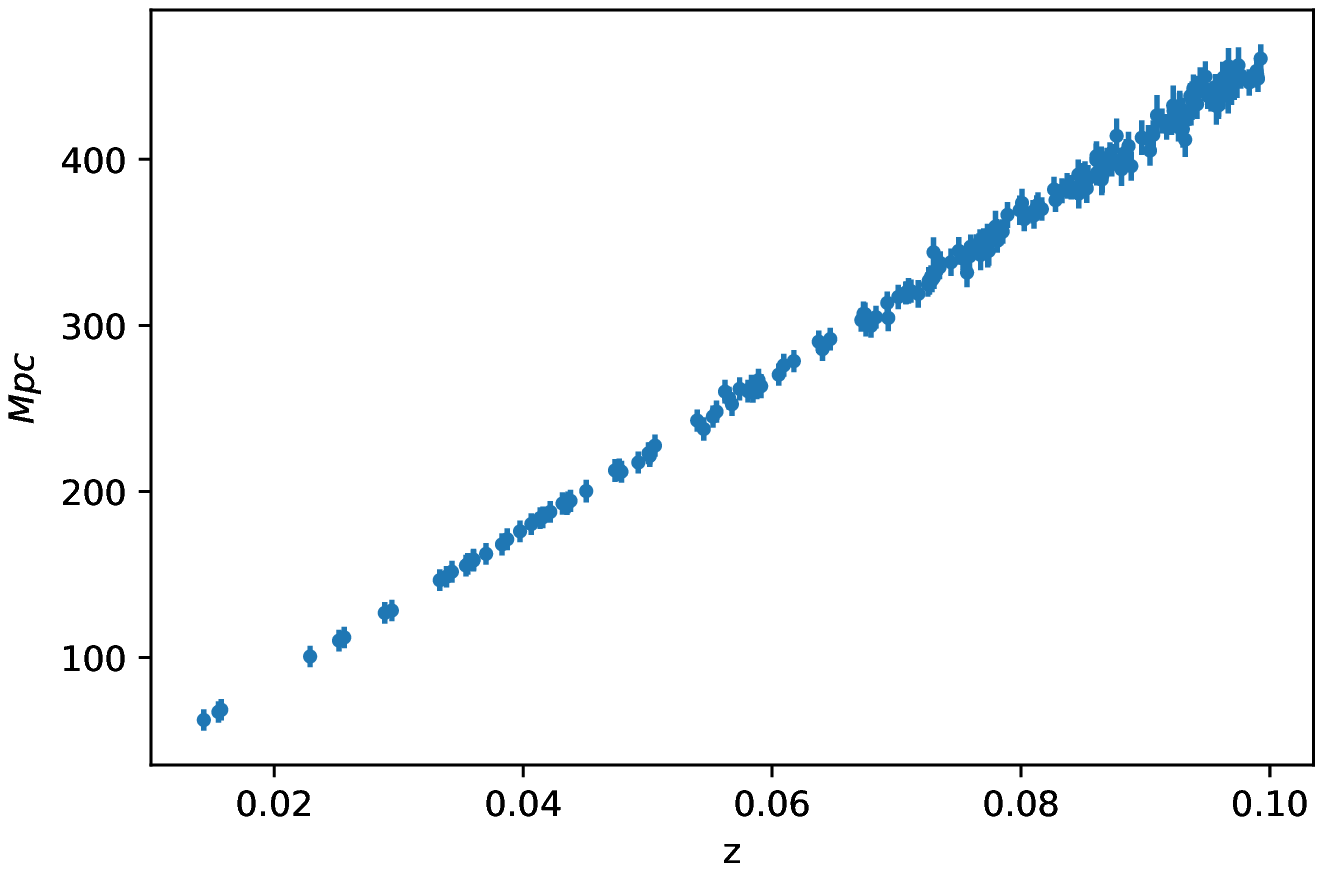}\includegraphics[width=0.6\hsize]{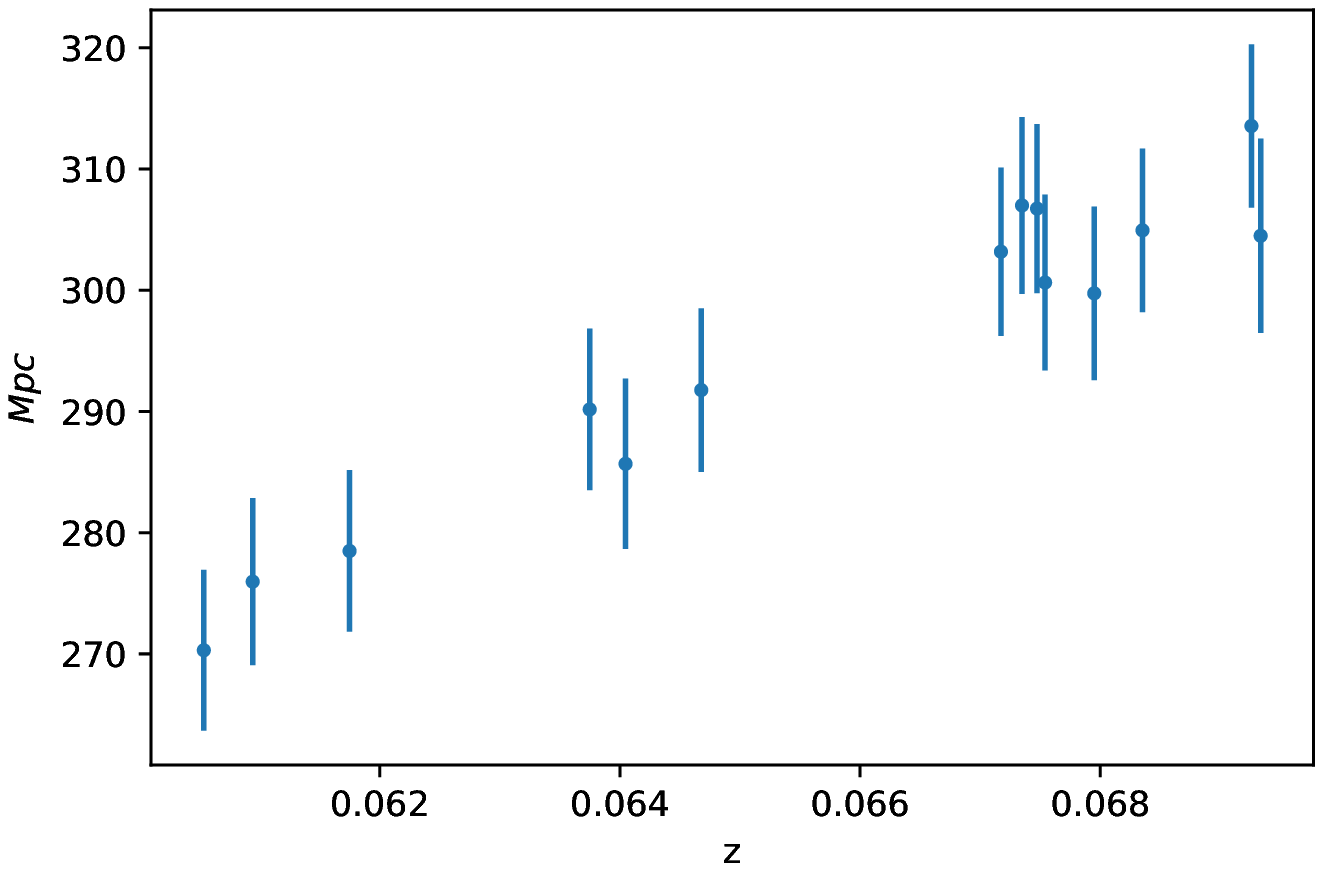}
\caption{The luminosity distance measurements from 200 low-redshift
GW events generated from inspiraling binary neutron stars (left
panel). Details of the measurements in the redshift range
$0.06~0.07$ are also shown for comparison (right panel).}
\end{figure*}

\section{Simulation and error estimation}

In this section we simulate GW events based on the Einstein
Telescope, the third generation of the ground-based GW detector. In
the simulation, the mass distribution of NS is chosen to be uniform
in the in interval of [1.4, 2.4] $M_{\bigodot}$. We adopt the
redshift distribution of the GW sources observed on Earth, which can
be written as \cite{Zhao11}
\begin{equation}
P(z)\propto \frac{4\pi D_c^2(z)R(z)}{H(z)(1+z)}, \label{equa:pz}
\end{equation}
where $H(z)$ is the Hubble parameter of the fiducial cosmological
model, $D_c(z)$ is the co-moving distance at redshift $z$, and
$R(z)$ represents the time evolution of the burst rate taken as
\cite{Schneider01}
\begin{equation}
R(z)=\begin{cases}
1+2z, & z\leq 1 \\
\frac{3}{4}(5-z), & 1<z<5 \\
0, & z\geq 5.
\end{cases}
\label{equa:rz}
\end{equation}

For the network of three independent ET interferometers, the
combined signal-to-noise ratio (SNR) of the GW waveform is
\begin{equation}
\rho=\sqrt{\sum\limits_{i=1}^{3}\left\langle
\mathcal{H}^{(i)},\mathcal{H}^{(i)}\right\rangle}. \label{euqa:rho}
\end{equation}
Here the inner product is defined as
\begin{equation}
\left\langle{a,b}\right\rangle=4\int_{f_{\rm lower}}^{f_{\rm
upper}}\frac{\tilde a(f)\tilde b^\ast(f)+\tilde a^\ast(f)\tilde
b(f)}{2}\frac{df}{S_h(f)}, \label{euqa:product}
\end{equation}
where $\tilde a(f)$ and $\tilde b(f)$ are the Fourier transforms of
the functions $a(t)$ and $b(t)$. $S_h(f)$ is the one-side noise
power spectral density (PSD) characterizing the performance of a GW
detector
\begin{equation}\label{eq11}
S_h(f)=S_0[x^{p1}+a_1x^{p2}+a_2f(x)]
\end{equation}
where $f(x)$ takes the form as
\begin{equation}\label{eq12}
f(x)=\frac{1+b_1x+b_2x^2+b_3x^3+b_4x^4+b_5x^5+b_6x^6}{1+c_1x+c_2x^2+c_3x^3+c_4x^4}
\end{equation}
with the definition of $x=f/200$ \cite{Zhao11}. The lower cutoff
frequency $f_{\rm lower}$ is fixed at 1 Hz. The upper cutoff
frequency, $f_{\rm upper}$, is decided by the last stable orbit
(LSO), $f_{\rm upper}=2f_{\rm LSO}$, where $f_{\rm
LSO}=1/(6^{3/2}2\pi M_{\rm obs})$ is the orbit frequency at the LSO,
and $\mathcal{M}_{\rm obs}=(1+z)\mathcal{M}_{\rm phys}$ is the
observed total mass. Meanwhile, the signal is identified as a GW
event only if the ET interferometers have a network SNR of
$\rho>8.0$ \cite{Cai17}.

Different sources of uncertainties are included in our simulation of
luminosity distance $D_L$. Firstly, in the standard framework of GW
data analysis, the information of relevant parameters
($\mathcal\theta$) are derived by fitting the frequent-domain wave
model $h(f;\mathbf{\theta}$) to the frequent-domain detector data
$d(f)$. If the noise is stationary and Gaussian, the likelihood
function $\mathcal L$ is defined as \cite{Li13}
\begin{equation}
\ln \mathcal{L}=-\frac{1}{2} \int_{0}^{\infty}df
\frac{|d(f)-h(f;\mathbf{\theta})|^{2}}{S_h(f)}
\end{equation}
where $d(f)$ and $S_h(f)$ are implicit functions of the calibration
parameters ($\mathbf{\lambda}$). Specially, considering the
difference between the detector's true calibration parameters
($\mathbf{\lambda_t}$) and the calibration parameters used to
produce the strain data from power fluctuations
($\mathbf{\lambda}$), the generated calibration error should be
included into the parameter estimation pipeline, described by the
parameter $\mathbf{\Delta \lambda}=
\mathbf{\lambda}-\mathbf{\lambda_t}$. In this analysis, we take the
calibration-induced error ($\sigma_{D_L}^{\rm cal}$) as one third of
the noise-induced error ($\sigma_{D_L}^{\rm noi}$), when the
parameter estimation is dominated by systematics induced by detector
noise \cite{Hall19}.

Secondly, when the error on luminosity distance is uncorrelated with
errors on the remaining GW parameters, the noise-induced error can
be estimated with Fisher matrix by \cite{Cutler09}
\begin{align}
\sigma_{D_L}^{\rm noi}\simeq \sqrt{\left\langle\frac{\partial
\mathcal H}{\partial D_L},\frac{\partial \mathcal H}{\partial
D_L}\right\rangle^{-1}},
\end{align}
It should be pointed out that in the ET era, we will be confronted
with a family of phenomenological waveforms that incorporates the
dynamics of the inspiral, merger, and ringdown phases of the
coalescence \cite{Husa16,Khan16}, while the inclusion of the merger
phase and ringdown phase may helpfully break the degeneracy between
the luminosity distance $D_L$ and inclination angle $\iota$
\cite{McWilliams12,Klein16}. In this paper, following the procedure
extensively applied in the literature \cite{Cai15,Cai17}, we focus
only on the inspiral phase of the GW signal, with the corresponding
instrumental error written as
\begin{equation}
\sigma_{D_L}^{\rm inst} = \sqrt{(\sigma_{D_L}^{\rm
noi})^2+(\sigma_{D_L}^{\rm cal})^2}
\end{equation}
where $\sigma_{D_L}^{\rm noi}$ is the noise-induced error that can
be estimated as $\sigma_{D_L}^{\rm noi}\simeq \frac{2D_L}{\rho}$
\cite{Cutler09,Li13}. Note that the maximal effect of the
inclination on the SNR is a factor of 2 (between $\iota=0^\circ$ and
$\iota=90^\circ$), for a conservative estimation of the correction
between $D_L$ and $\iota$.

Thirdly, following the strategy described by \cite{Cai17}, weak
lensing has been estimated as a major source of error on $D_L(z)$
for standard sirens. For the ET we estimate the uncertainty from
weak lensing according to the fitting formula of
$\sigma_{D_L}^{lens}/D_L=0.05z$ \cite{Zhao11}. Therefore, the
distance precision per GW is taken to be
\begin{align}
\sigma_{D_L}^{\rm sta}&~~=\sqrt{(\sigma_{D_L}^{\rm
inst})^2+(\sigma_{D_L}^{\rm lens})^2}. \label{sigmadl}
\end{align}

Finally, precise redshift measurements are the crucial point of our
idea. We consider two cases: In the with-counterpart case, with the
observation of the EM counter parts, the redshift of a GW event can
be determined. We assume that the EM counterpart is close enough to
its host galaxy that the host can be unambiguously identified, and
we can measure its sky position and redshift. In the previous works
\cite{Cai15,Cai17,Qi19a,Qi19b}, the uncertainty of the redshift
measurement is always ignored, because it is ignorable compared to
the uncertainty of the luminosity distance. However, in the case of
local universe, the redshift uncertainty caused by the uncertainty
of peculiar velocity should be taken into account. Throughout this
paper, we take the redshift $z$ to be the
peculiar-velocity-corrected redshift (i.e., the redshift of the
source located in the Hubble flow) and apply two different cases.
Following the procedure performed in the recent analysis
\cite{Chen18}, a standard deviation of $c\sigma_z=$200 km/s is
assumed for each BNS and BH-NS system (with a direct EM
counterpart), which is a typical uncertainty for the peculiar
velocity correction well consistent with the peculiar velocity
measurement of NGC 4993 (the host galaxy of GW170817)
\cite{Nicolaou19}. Note that the recent analysis of BH-NS mergers
has discussed the possibility that the neutron star can be tidally
disrupted and emit electromagnetic radiation, depending on the mass
ratio and the black hole spin \cite{Vitale18}. For the latter case
in which the host galaxy of a GW event can not be identified (BBH),
the redshift comes from a statistical analysis over a catalogue of
potential host galaxies, which will be discussed later.

Now the final key question required to be answered is: how many
low-redshift GW events can be detected per year for the ET? Focusing
on the GW sources caused by binary merger of neutron stars (with
detectable EM counterpart measurable source redshift), it is
revealed that the third generation ground-based GW detector can
detect up to $10^3$ - $10^7$ events, with the upper detection limit
of $z\sim 2.0$ \cite{ET}. Following our detailed calculation that
only 0.1\% of the total GW events will be located in the redshift
range of $[0,0.1]$, one may expect that $1$ - $10^4$ low-redshift
events could be used in our analysis. In addition, recent analysis
revealed that the five-detector network including LIGO, Virgo, KAGRA
and LIGO-India plans to detect $\sim40$ events per year, if the
designed sensitivity of the network could be achieved in the
future~\cite{Chen18}. Therefore, assuming the luminosity distance
measurements obey the Gaussian distribution, we simulate 200 GW
events of BNS merging used for statistical analysis in the next
section, the redshift distribution of which is shown in Fig.~1.

We summarize the main route of our method as follows:

\begin{itemize}
\item Simulate 200 GW events according to the redshift distribution
in Eq.~(\ref{equa:pz}). The angles describing the position of each
BNS system are randomly sampled within the interval of $\theta\in
[0, \pi)$ and $\phi\in [0, 2\pi)$.

\item Calculate the luminosity distance $D_L(z)$ according to Eq.
(2)-(3). Randomly sample the mass of neutron star within
$[1.4,2.4]M_{\bigodot}$. Evaluate the signal-to-noise ratio (SNR)
and the error $\sigma_{D_L}$ when the SNR of the detector network
reaches above 8.

\item For a confirmed GW event, the statistical error of luminosity
distance $\sigma_{D_L}{\rm sta}$ can be figured out through
Eq.~(17). Moreover, considering the effect of the peculiar velocity
of the host galaxy $v_{\rm pec}$, we also add the systematical
uncertainty of observed redshift $z_{\rm obs}$ to project
uncertainties $\sigma_{D_L}^{\rm sys}$ onto the final uncertainty of
distance estimation (Eq. (3)). We therefore take the total
uncertainty on the luminosity distance as $\sigma_{D_L}^{\rm
tot}=\sqrt{(\sigma_{D_L}^{\rm sta})^2+(\sigma_{D_L}^{\rm sys})^2}$.
The observed luminosity distance $D_L$ follows a Gaussian
distribution whose mean is $D_{L}^{fid}$ and variance is
$\sigma_{D_L}^{\rm tot}$, i.e.,
$D_L\sim\mathcal{N}(D_{L}^{fid},\sigma_{D_L}^{\rm tot})$.

\end{itemize}

\begin{figure*}\label{fig2}
\centering
\includegraphics[width=0.6\hsize]{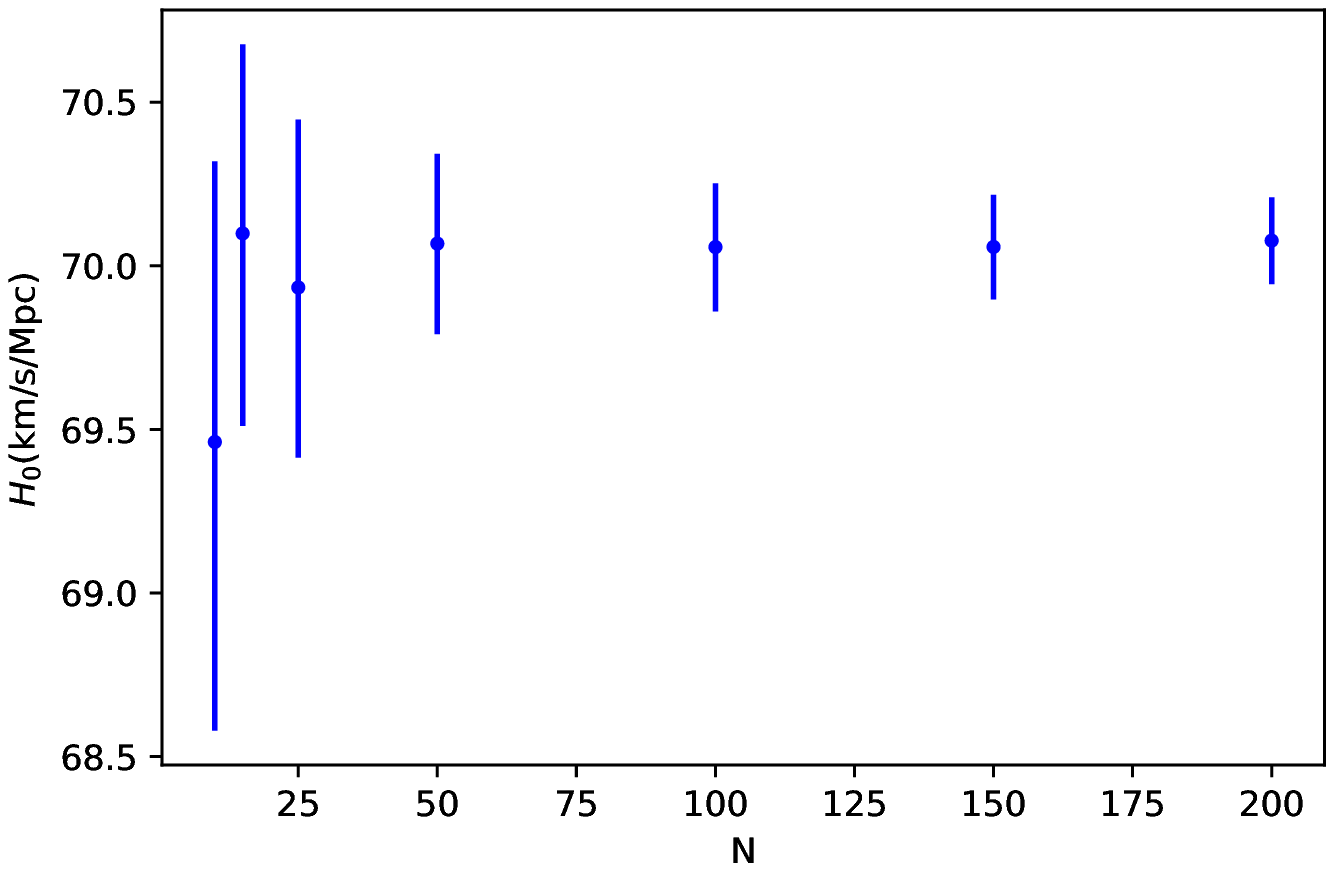}\includegraphics[width=0.6\hsize]{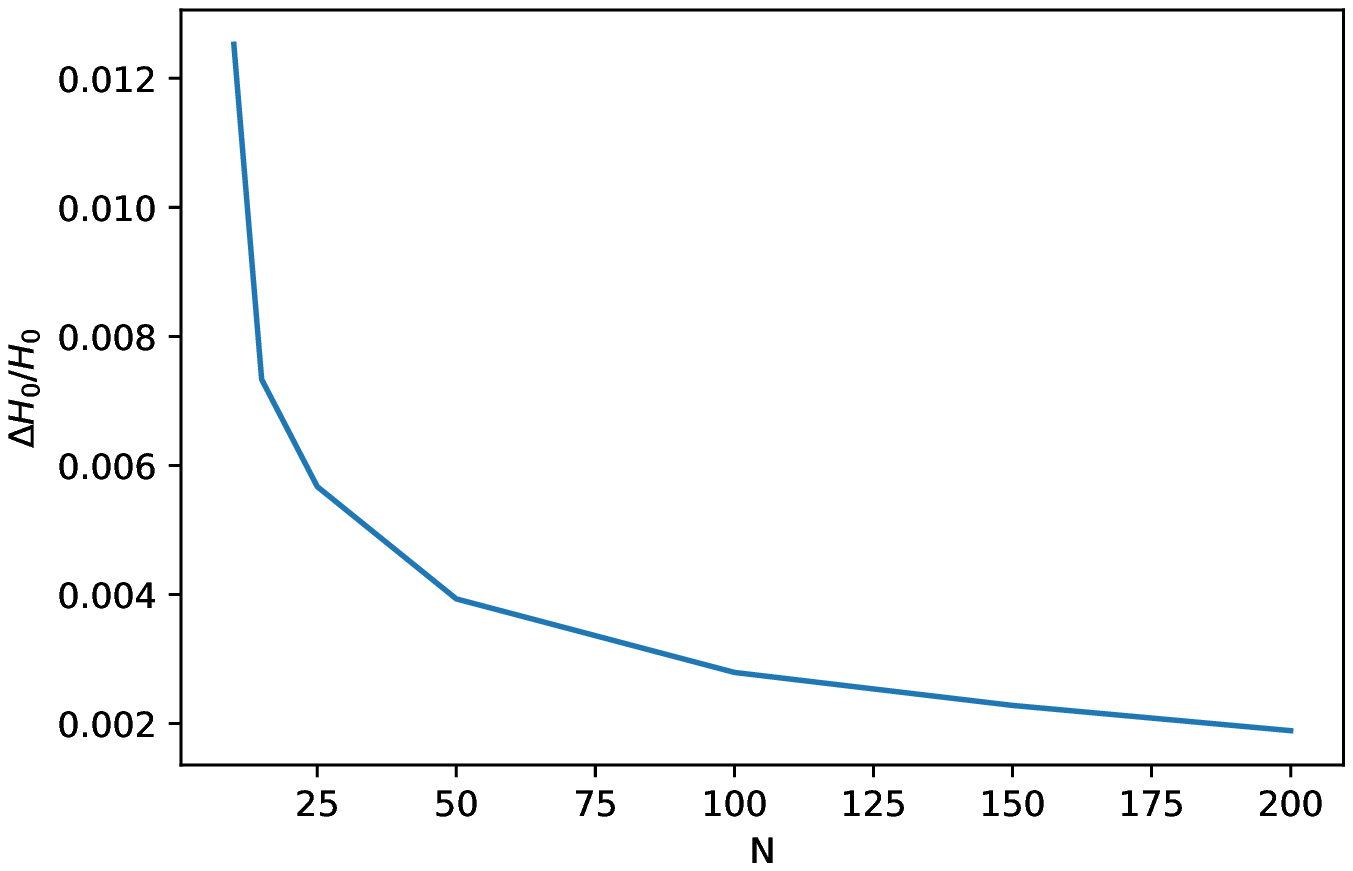}
\caption{Left: Inferred Hubble constant as a function of the number
of GW events (BNS mergers with EM counterparts). Right: The
corresponding precision of the Hubble constant constraints for a
variable number of GW events (BNS mergers with EM counterparts).}
\end{figure*}

\section{Results and discussions}

In order to place constraints on the Hubble constant with MCMC
method, the likelihood estimator is determined by $\chi^2$
statistics
\begin{equation}\label{eq19}
\chi^2=\sum_{i=1}^{N}\left(\frac{D_{L}^{obs}(z_i)-D^{th}_{L}(z_i)}{\sigma_{D_{L,i}}}\right)^2
\end{equation}
where $N$ denotes the number of data sets, $D^{th}_{L}$ is the
predicted luminosity distance value in the Hubble law and
$D^{obs}_{L}$ is the measured value with a uncertainty of
$\sigma_{D_{L}}$ in the simulated data.

Fig.~2 shows the precision of the curvature parameter assessment as
a function of GW sample size for future ET detector, where weighted
means and corresponding standard deviations are illustrated for
comparison. Projected fractional error for the standard siren $H_0$
measurement for BNSs is also shown. One can see that, even with
about 20 well-observed GW events due to BNS mergers one can expect
the Hubble constant to be estimated with the precision of $\delta
H_0\sim 1\%$ (at the 1$\sigma$ level), if it is possible to
independently measure a unique redshift for all BNS events. More
importantly, we find that in this counterpart case, the fractional
$H_0$ uncertainty will be proportional to $1/\sqrt{N}$, where $N$ is
the number of BNS mergers detected by the ET. Still, there are
several remarks that remain to be clarified as follows.

Firstly, in the above analysis we assume that EM counterparts are
detectable for all BNS systems. For example, as GW170817
demonstrated, for the merger of a BNS system it is possible to
identify a kilonova counterpart independently of the short $\gamma$
ray burst (SGRB). These EM counter parts can help us in locating the
events in the sky and identifying the host galaxy of the event
\cite{Nakar07}, and the locating ability can be improved with the
use of H.E.S.S.Imaging Air Cherenkov Telescopes(IACTs)
\cite{H.E.S.S19}. However, from observational point of view we don't
expect to observe EM counter parts for all GW events.For example,
the optical counter parts, kilo-nova, are too dim to be observed in
large luminosity distance. More specifically, SGRBs are strongly
beamed phenomena which carry a great deal of energy and we can only
detect them when they are almost face-on, which will significantly
decline the number of detectable GW events with SGRB
\cite{Rezzolla11}. Following the recent discussion of GW170817, the
eject matter from a merging BNS system can cause a secondary
radiation which might power the radiation for a longer time and a
wider radiation angle. This suggest that only $10\%$ of the 200 BNS
events detected by ET will be available with the EM counterparts
such as SGRB \cite{Chen18}. Therefore, after about 20
gravitational-wave standard sirens, the fractional uncertainty on
$H_0$ could still reach 1\% (at the 1$\sigma$ level) by the end of
five years of ET at design sensitivity, sufficient to arbitrate the
current tension between local and high-$z$ measurements of $H_0$.

Secondly, besides BNS with EM counterparts, there are other types of
GW sources that can contribute in providing precise determination of
the Hubble constant. Theoretically, ET could also detect a large
number of GW signals for black hole - neutron star (BH-NS) merger
systems and binary black hole (BBH) merger systems. These two types
of GW sources are also of concern to our investigation in this
paper. While we do not expect EM counter parts from BBH systems, we
do want to observe the EM counter parts from BN-BH systems, and the
optical luminosity depends on the property of the neutron
star\cite{Barbieri19}. For the former type, the electromagnetic (EM)
signals are emitted during the merger processes, allowing us to
determine the redshift of sources. In the framework of ET project,
the expected detect rate of BNS and BHNS are in the same order of
$10^3$ to $10^7$ \cite{ET}. Therefore, it is reasonable to assume
that with the observation of 20 BNS merger with EM counter parts,
one can detect 20 BHNS merger with their EM counterparts as well.
For the latter case where a unique counterpart cannot be identified
for a BBH merger, it is possible to carry out a measurement of the
Hubble constant using the statistical approach, i.e., there are
still other methods to identify their host galaxies, in a
statistical way which might not be that accurate \cite{ET}. More
specifically, we will apply the methodology proposed in
\cite{Chen18}, in which a galaxy catalogue is used to describe all
potential host galaxies in the case that the EM counterpart is
absent. The simulated galaxy catalogue is constructed by
distributing galaxies uniformly in the co-moving volume of 10000
Mpc$^{3}$ with a number density of 0.02 Mpc$^{-3}$, each of which
has the same probability to be the host galaxy of the GW event.
Given the detailed calculation presented in \cite{Chen18}, after two
years of full operation, the LIGO and Virgo network is expected to
detect $\sim 16$ BBH events (with well estimation of statistical
redshifts) in the local universe (located in the volume of 10000
Mpc$^{3}$), which will lead to a $10\%$ of Hubble constant
measurement. Therefore, it is reasonable to simulate 30 BBH systems
with redshift determination in the framework of ET configurations
(in the simulation, the mass distribution of BH is chosen to be
uniform in the in interval of [3,10] $M_{\odot}$
\cite{Qi19a,Qi19b}). A representative result is illustrated in
Fig.~3, in order to compare the $H_0$ constraints with different
types of gravitational-wave events. The left panel is obtained using
BNS GW sirens, while the middle panel is obtained using BHNS sirens,
which is compared in the right panel with BBH GW sirens,
respectively. As was noted in the previous work based on the LIGO
and Virgo network \cite{Chen18}, the error bars will be greatly
reduced when different types of GW events are included. Meanwhile,
benefit from a larger total mass and smaller merger frequency than
BNS, one could also expect a larger SNR for a specific BHNS event,
which will greatly contribute to the distance measurements and thus
the standard siren $H_0$ constraint \cite{Vitale18}. However,
constraints from BBH systems without counterparts are inferior, due
to the larger number of potential host galaxies compared with other
two types of GW events \cite{Chen18}.

\begin{figure}\label{fig3}
\centering
\includegraphics[width=0.6\hsize]{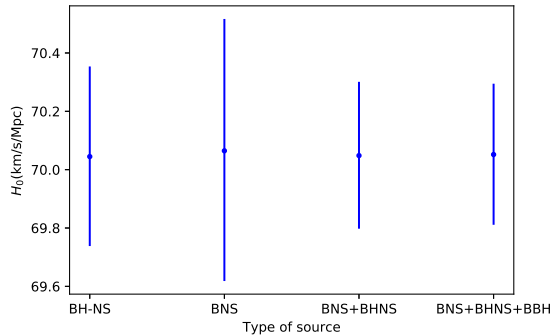}
\caption{Constraints on the Hubble constant with three types of GW
sources, 20 BNS with EM counterparts, 20 BH-NS with EM counterparts,
and 30 BBH without EM counterparts but with well-estimated
statistical redshifts.}
\end{figure}

Now it is worthwhile to compare our forecast results with some
previous $H_0$ tests fitting the Hubble constant in different
cosmological models, based on the ``$D_L - z$" relation at higher
redshifts in the GW domain. Using the information of luminosity
distances and redshifted chirp masses for a catalog of BNSs detected
by an advanced era network, \cite{Taylor12} studied a technique to
obtain constraints on the Hubble constant and NS mass-distribution
parameters simultaneously. It was found that $H_0$ could be
estimated at the precision of $10\%$ with $\sim 100$ such kind of GW
events. Meanwhile, in the framework of a range of ground-based
detector networks, \cite{Nissanke10} examined how well distances
(and thus cosmological parameters) can be measured from BNSs with
electromagnetic counterparts (such as the associated SGRB). The
analysis results revealed that $H_0$ could be measured with a
fractional error of $\sim 13\%$ with 4 GW-SGRB events, which could
be improved to $\sim 5\%$ with 15 events detected by the advanced
LIGO-Virgo detector network. Focusing on constraint ability of the
third-generation gravitational wave detector (the Einstein
Telescope), the recent analysis showed that with the simulated data
of 500 standard sirens, one can constrain the Hubble constant with
an accuracy comparable to the most recent Planck results
\cite{Cai17}. By considering our results and those from
\cite{Taylor12,Nissanke10,Cai17}, our results show that strong
constraints on the Hubble constant can be obtained in a
cosmological-model-independent fashion. Such conclusion agrees very
well with that obtained in the framework of LIGO-Virgo detector
network \cite{Chen18}.

Measuring the Hubble constant ($H_0$) independent of CMB
observations is one of the most important complementary probes for
understanding the nature of the Universe. Therefore, the fractional
uncertainty on $H_0$ will reach 1\% by the end of five years of ET
at design sensitivity, which furthermore strengthens the probative
power of the third generation ground-based GW detectors to inspire
new observing programs or theoretical work in the moderate future.
Finally, one should note that, in order to achieve this goal,
dedicated observations of the sky position of each host galaxy (that
is, with negligible measurement error) would be necessary. Although
the GW distance posterior changes slowly over the sky and therefore
is not sensitive to the precise location of the counterpart,
obtaining such measurements for a sample of different types of GW
events would require substantial follow-up efforts, which can lead
to significant improvements in the distance, and hence $H_0$
measurements. We also hope future observational data such as
strongly lensed gravitational waves (GWs) from compact binary
coalescence and their electromagnetic (EM) counterparts systems
\cite{Liao17,Cao19a}, precise measurements of the Hubble parameter
obtained by cosmic chronometer and radial BAO size methods
\cite{Ding15,Zheng16,Qi18}, and VLBI observations of compact radio
quasars with higher sensitivity and angular resolution
\cite{Cao18,Cao19b,Xu18,Ma19} may improve remarkably the constraints
on this key cosmological parameter.

\section*{Acknowledgments}

We are grateful to Jingzhao Qi for helpful discussions. This work
was supported by National Key R\&D Program of China No.
2017YFA0402600, the National Natural Science Foundation of China
under Grants Nos. 11690023, 11373014, and 11633001, the Strategic
Priority Research Program of the Chinese Academy of Sciences, Grant
No. XDB23000000, the Interdiscipline Research Funds of Beijing
Normal University, and the Opening Project of Key Laboratory of
Computational Astrophysics, National Astronomical Observatories,
Chinese Academy of Sciences.


\begin{thebibliography}{99}

\bibitem{Weinberg13} Weinberg, D. H., et al. Phys. Rep., 530, 87 (2013)
\bibitem{Ade16} Ade, P. A. R., et al. (Plank Collaboration), A\&A, 594, A13 (2016)
\bibitem{Cao17} Cao, S., et al. A\&A, 606, A15 (2017)
\bibitem{Cao11}  Cao, S., Liang N., \&  Zhu, Z.-H. MNRAS, 416, 1099 (2011)
\bibitem{Cao13} Cao, S. \& Liang, N. IJMPD, 22, 1350082 (2013)
\bibitem{Cao15} Cao, S., et al. IJTP, 54, 1492 (2015)
\bibitem{Chen15} Chen, Y., et al. JCAP, 02, 010 (2015)
\bibitem{Pan15} Pan, Y., et al. ApJ, 808, 78 (2015)
\bibitem{Freedman17} Freedman, W. L. Nature Astronomy, 1, 0121 (2017)
\bibitem{Wong19} Wong, K. C., et al. arXiv:1907.04869
\bibitem{Pan16} Pan, Y., et al. IJMPD, 25, 1650003 (2016)
\bibitem{Riess19} Riess, A. G., et al. ApJ, 876, 85 (2019)
\bibitem{Schutz86} Schutz, B. F. Nature, 323, 310 (1986)
\bibitem{Abbott16}  Abbott, B. P., et al. (LIGO Scientific and Virgo Collaborations), PPL, 116, 061102 (2016)
\bibitem{Abbott17} Abbott, B., et al. (LIGO Scientific and Virgo Collaborations), 2017, Nature, 551, 85
\bibitem{Holz05}  Holz, D. E. \& Hughes, S. A., ApJ, 629, 15 (2005)
\bibitem{MacLeod08} MacLeod, C. L. \&  Hogan, C. J., Phys. Rev., 77, 043512 (2008)
\bibitem{Sathyaprakash10} Sathyaprakash, B. S., et al. CQG, 27, 215006 (2010)
\bibitem{Zhao11} Zhao, W., et al. PRD, 83, 023005 (2011)
\bibitem{Cai15} Cai, R.-G., et al. arXiv:1509.06283
\bibitem{Qi19a} Qi, J. Z., et al. PRD, 99, 063507 (2019)
\bibitem{Qi19b} Qi, J. Z., et al. 2019, PDU, 26, 100338 (2019)
\bibitem{Nissanke10} Nissanke, S., et al. ApJ, 725, 496 (2010)
\bibitem{Taylor12}  Taylor, S. R., Gair, J. R., \& Mandel, I. PRD, 85, 023535 (2012)
\bibitem{Cai17} Cai, R.-G. \& Yang, T. PRD, 95, 044024 (2017)
\bibitem{Chen18} Chen, H.-Y., et al. Nature, 562, 545 (2018)
\bibitem{ET} The Einstein Telescope Project, https://www.et-gw.eu/et/
\bibitem{Hubble29} Hubble, E. PNAS, 15, 168 (1929)
\bibitem{Harrison93} Harrison, E. ApJ, 403, 28 (1993)
\bibitem{Hogg99} Hogg, D. W. arXiv:9905116v4
\bibitem{Carroll07} Carroll, B. W. \& Ostlie, D. A., An Introduction to Mordern Astrophysics (2nd Edition), p1056, Cambridge University Press (2007)
\bibitem{Cai18} Cai, R.-G., et al. PRD, 97, 103005 (2018)
\bibitem{Schneider01} Schneider, R., et al. MNRAS, 324, 797 (2001)
\bibitem{Li13} von Toussaint, U. Rev. Mod. Phys. 83, 943 (2011)
\bibitem{Hall19} Hall, E. D., et al. arXiv:1712.09719
\bibitem{Cutler09} Cutler, C. \& Holz, D. E. PRD, 80, 104009 (2009)
\bibitem{Husa16} Husa, S., et al. PRD, 93, 044006 (2016)
\bibitem{Khan16} Khan, S., et al. PRD, 93, 044007 (2016)
\bibitem{McWilliams12} McWilliams, S. T., Lang, R. N., Baker, J. G., \& Thorpe, J. I. PRD, 84, 064003 (2012)
\bibitem{Klein16} Klein, A., et al. PRD, 93, 024003 (2016)
\bibitem{Nicolaou19}Nicolaou, C., Lahav, O., Lemos, P., Hartley, W., and Braden, J. arXiv:1909.09609
\bibitem{Vitale18} Vitale, S. \& Chen, H. Y. PRL, 112, 251101 (2018)
\bibitem{Nakar07} Nakar, E. Phys. Rep. 442, 166 (2007)
\bibitem{H.E.S.S19} Monica Seglar-Arroyo, M., et al. [The H.E.S.S Collaboration], arXiv:1908.08822v1
\bibitem{Rezzolla11} Rezzolla, L., et al. ApJL, 732, L6 (2011)
\bibitem{Barbieri19} Barbieri, C., et al. arXiv:1908.08822v1
\bibitem{Liao17} Liao, K., et al. Nature Communications, 8, 1148 (2017)
\bibitem{Cao19a} Cao, S., et al. NatSR, 9, 11608 (2019)
\bibitem{Ding15} Ding, X., et al. ApJL, 803, L22 (2015)
\bibitem{Zheng16} Zheng, X., et al. ApJ, 825, 17 (2016)
\bibitem{Qi18} J.-Z. Qi, et al. RAA, 18, 66 (2018)
\bibitem{Cao18} Cao, S., et al. EPJC, 78, 749 (2018)
\bibitem{Cao19b} Cao, S., et al. PDU, 24, 100274 (2019)
\bibitem{Xu18}  Xu, T. P., et al. JCAP, 06, 042 (2018)
\bibitem{Ma19}  Ma, Y. B., et al. EPJC, 79, 121 (2019)














\end{thebibliography}
\end{document}